\documentclass{./jpsj-suppl}
\usepackage{txfonts} 

\title{Properties and Significance of the Surface Dipole Mode}

\author{P. \textsc{Papakonstantinou}}

\inst{Rare Isotope Science Project, Institute for Basic Science, Daejeon 305-811, Republic of Korea}

\email{ppapakon@ibs.re.kr}

\recdate{\today}

\abst{%
A strong isoscalar dipole resonance is known to be excited in a variety of nuclei, including isospin symmetric ones, at approximately $6-7$~MeV. 
A series of theoretical studies and accumulating experimental evidence  
support an interpretation of the above dipole resonance as an elementary surface vibration. 
Obviously, such a mode is potentially as interesting as any 
collective excitation for a variety of reasons. 
In addition, though, it is found to account for the observed isoscalar segment of pygmy dipole strength. 
As discussed here, this has important implications for pygmy-strength interpretations and searches for genuine neutron-skin oscillations. }  

\kword{surface dipole mode,low-energy dipole strength,pygmy resonances}

\begin{document}
\maketitle

\section{Introduction}

Low-energy collective nuclear excitations represent fundamental nuclear properties~\cite{RoweWood}. 
Rotational modes arranged in characteristic bands signify, quite obviously, deformation. 
Other low-energy collective modes are interpreted as surface vibrations in the collective model and are confirmed as such microscopically, within the random-phase approximation (RPA) and extensions tehreof. 
In ordinary nuclei they are of predominately isoscalar character. 
The best studied vibrational modes are the quadrupole and octupole ones, 
predominately represented by a strong, well-defined low-energy transition. 
The properties of the latter reveal magicity or shape transitions. 
Hexadecapole and coupled phonons lie at somewhat higher energies and are typically more fragmented, marking the limits of harmonicity.

Could there be such a thing as a surface {\em dipole} vibration? 
A surface dipole mode--an isoscalar one--is more difficult to visualize, because, to lowest order, the isoscalar dipole field 
induces merely a translation of the nucleus as a whole. 
Nonetheless, it is possible to imagine a dipole oscillation of a surface layer against a nuclear core. 
Precisely such a picture has been used to describe the neutron-skin (or valence-neutron) mode; 
an {\em isoscalar} mode would involve a surface layer which does not contain only neutrons, but both species of nucleons. 
Rather than being simply surface-peaked, the transition density 
whould need to have a node to ensure translational invariance--the inner peak being associated with the core's translation. 
The node would result in low $E1$ strength. 
 
In fact, a strong low-energy {\em isoscalar dipole} excitation has been observed in a variety of stable nuclei, including isospin symmetric ones, since decades--cf. the results summarized in Ref.~\cite{Hav}, as well as, more recently, experimental reports on the isospin content of pygmy dipole states~\cite{SAZ2013,Der2013,Der2014,Cre2014} 
(data compiled in Fig.~\ref{Fig:IS-LED}(a)). 
We shall refer to this phenomenon as the isoscalar low-energy dipole (IS-LED) mode or strength. 
Beyond the RPA results summarized next, the IS-LED modes have received limited theoretical attention as such (with likely exceptions in Refs.~\cite{Bas2008,Urb2012}). 
In some studies focused on pygmy strength, the IS-LED strength is attributed to neutron-skin oscillations. 
There are two serious problems with such an interpretation: 
1) microscopic models which associate the IS-LED strength with neutron excess severely overestimate the isovector (or $E1$) strength~\cite{PVK2007,Cha1994}  
and 
2) a neutron-skin vibration can obviously not account for the strong isoscalar modes observed in $N=Z$ nuclei. 

The RPA calculations and related data surveyed here support an interpretation of the observed low-energy dipole transitions as collective surface vibrations.  
Such vibrations can therefore 
account for the isoscalar segment of the pygmy dipole strength--with important implications in related studies. 
Beyond pygmy-strength studies, it shall be pointed out that such collective vibrations deserve interest for their own sake, for validating theoretical models, 
as well as 
for possibly influencing the outcome of nuclear reactions. 

\section{Experimental evidence and case studies} 
\begin{figure}[t]
\includegraphics[angle=-90,width=80mm]{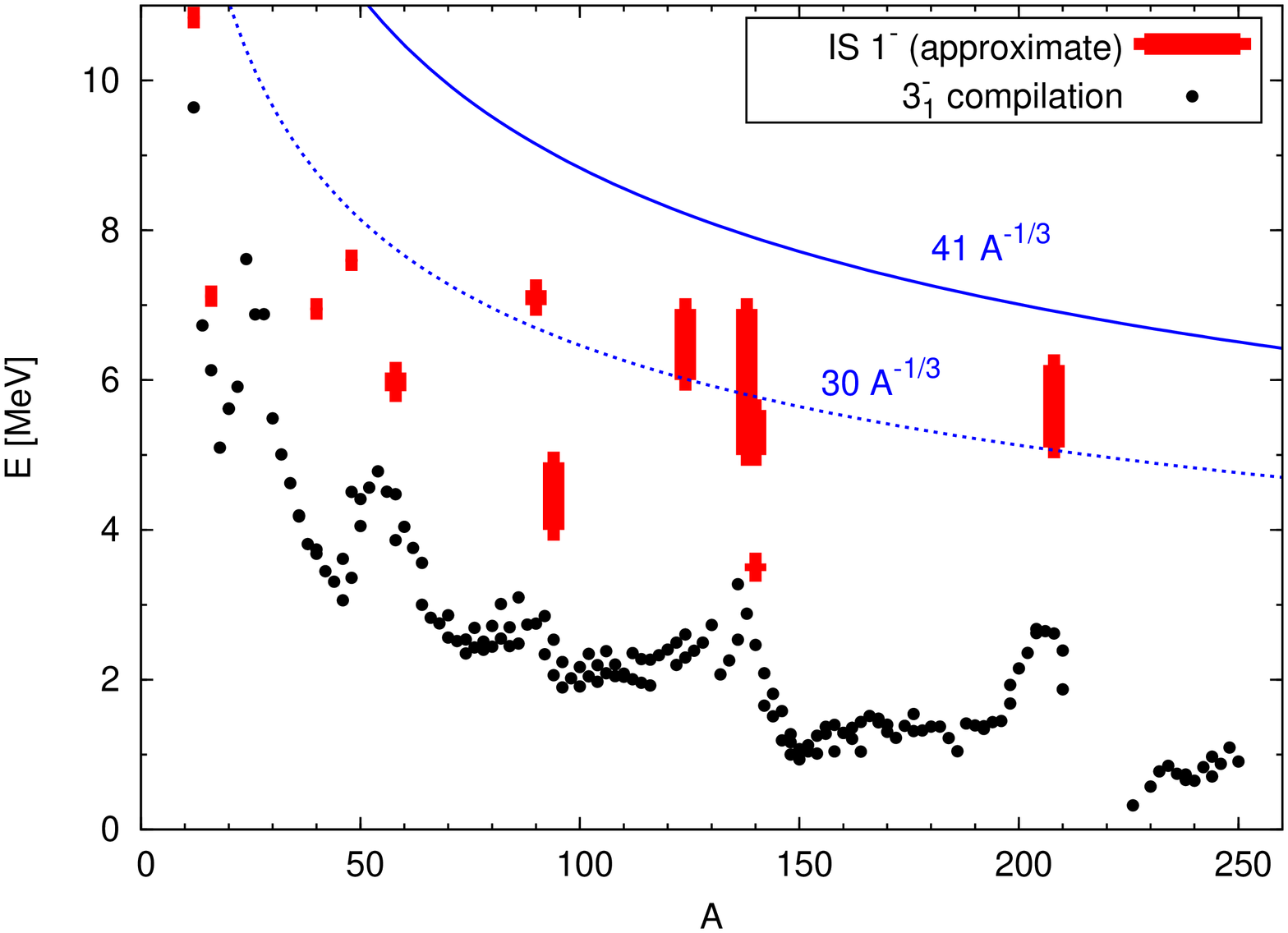} \,\, 
\includegraphics[angle=-90,width=60mm]{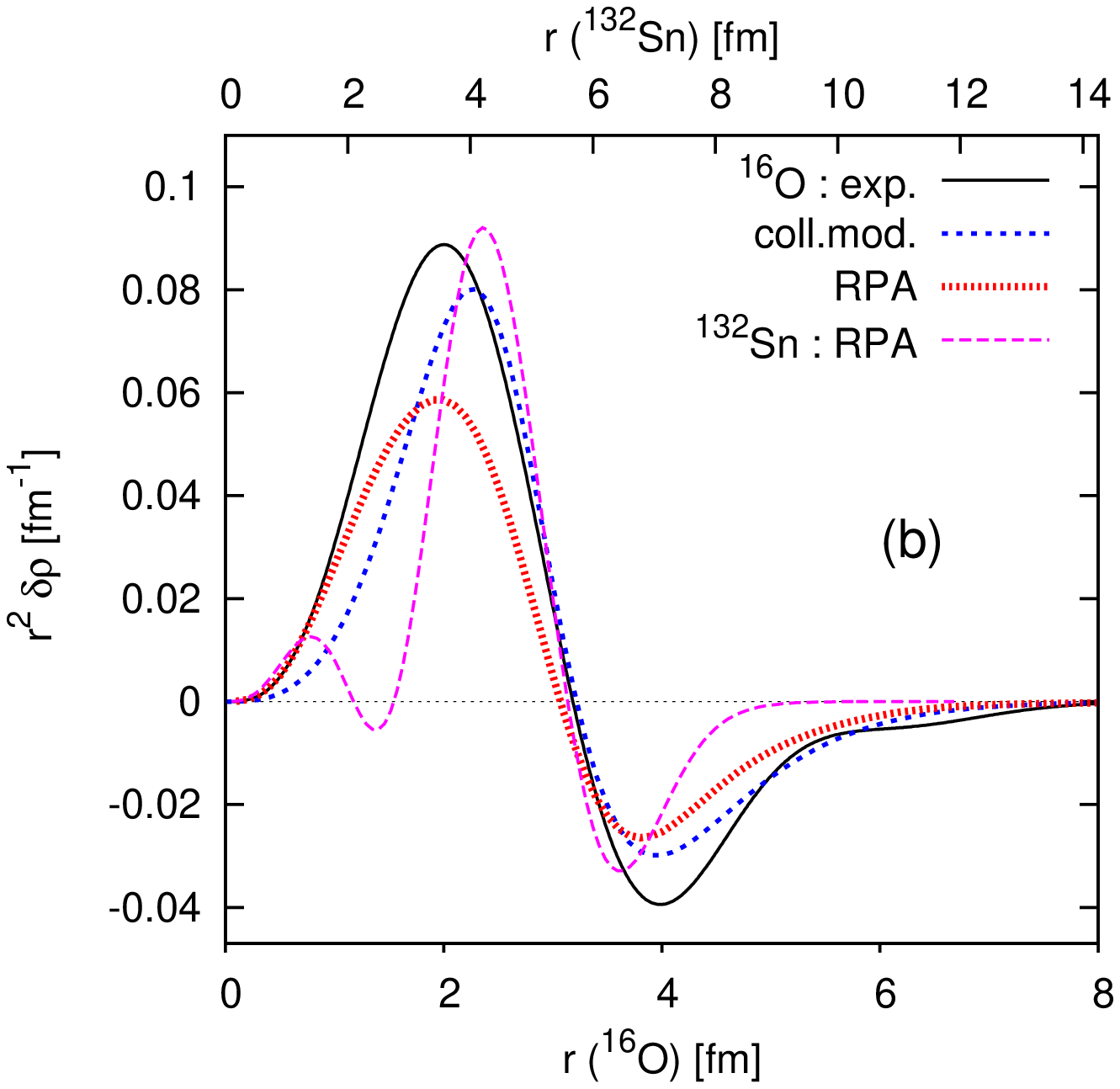}
\caption{(a) The energy (approximately depicted) of the observed low-energy isoscalar $1^-$ transitions in 
$^{12}$C~\cite{Ajz1990}, 
$^{16}$O~\cite{HaD1981,Ajz1986}, 
$^{40}$Ca~\cite{Hav,Poe1992}, 
$^{48}$Ca~\cite{Der2014}, 
$^{58}$Ni, 
$^{90}$Zr~\cite{Hav,Poe1992}, 
$^{94}$Mo\cite{Der2013}, 
$^{124}$Sn, 
$^{138}$Ba, 
$^{140}$Ce~\cite{SAZ2013}, 
$^{208}$Pb~\cite{Hav,Cre2014} in comparison with the energy of the first $3^-$ state~\cite{ADNDToct}, 
an average empirical energy of low-lying $3^-$ states, $30A^{-1/3}$~MeV~\cite{Hav}, and the empirical $1\hbar\omega$ value, 41$A^{-1/3}$. 
(b) 
The charge transition density of the IS-LED in $^{16}$O as extracted via electron scattering~\cite{But1986}, 
and the point-proton transition density  
as predicted: by RPA with the Gogny D1S interaction; 
and within a collective model~\cite{Hav} for the reported excitation energy and portion of the energy-weighted sum rule (assuming here an equilibrium density profile of Woods-Saxon type with $\rho_0=0.16$~fm$^{-3}$, $R=2.634$~fm and $a=0.465$~fm, such that the point-charge mean-square radius (r.m.s.) $R_{{\mathrm r.m.s.}^{16}\mathrm{O}}\approx 2.7$~fm). 
The RPA result for $^{132}$Sn is also shown, for the coordinate shown on the top axis $r(^{132}$Sn$)= r(^{16}$O$)\times R_{{\mathrm r.m.s.}^{132}\mathrm{Sn}}/R_{{\mathrm r.m.s.}^{16}\mathrm{O}}$ (point-proton r.m.s. radii as calculated with the Gogny D1S interaction). 
}
\label{Fig:IS-LED}
\end{figure}

Compiling the experimental results for 
$^{12}$C~\cite{Ajz1990}, 
$^{16}$O~\cite{HaD1981,Ajz1986}, 
$^{40}$Ca~\cite{Hav,Poe1992}, 
$^{48}$Ca~\cite{Der2014}, 
$^{58}$Ni, 
$^{90}$Zr~\cite{Hav,Poe1992}, 
$^{94}$Mo\cite{Der2013}, 
$^{124}$Sn, 
$^{138}$Ba, 
$^{140}$Ce~\cite{SAZ2013}, 
$^{208}$Pb~\cite{Hav,Cre2014}--see Fig.~\ref{Fig:IS-LED}(a)--we readily observe the following: 
The excitation energy of the IS-LED is typically $6-7$~MeV--somewhat higher for the lightest nuclei
or lower in open-shell nuclei. 
(In very light nuclei such as $^{12}$C a simplistic vibrational or shell-model picture may of course not apply.) 
It lies 
above the first octupole state, which is also an odd-parity $1\hbar\omega$ mode, 
but below the empirical $1\hbar\omega$ value of $41A^{-1/3}$~MeV. 
Whenever evaluated, the portion of the isoscalar energy-weighted sum rule exhausted by low-lying dipole transitions has been found to amount to a few percentage points~\cite{Hav,HaD1981,Poe1992,Der2014,Cre2014}--a huge amount for states at such low energies. 

Experimentally the IS-LED strength is often found fragmented. In many cases one state dominates, especially in magic nuclei~\cite{HaD1981,Poe1992,Hav,Der2014,Cre2014}. 
RPA and quasiparticle RPA (QRPA) calculations, in a harmonic-oscillator basis, of the IS-LED modes in $N=Z$ nuclei \cite{PPR2011} and in Ca~\cite{PHP2012,Der2014} and Sn~\cite{PHP2014} isotopes 
overestimate the excitation energy by about 3~MeV~\cite{PPR2011,PHP2012,Der2014,PHP2014}, 
which is not surprising for this type of calculations, 
but reproduce the isoscalar strength rather well. 
The Gogny D1S interaction roughly reproduces also the $E1$ strength and is used in the following discussion. 

\subsection{$N=Z$ nuclei: Structure of the IS-LED vibration} 
 
In the $N=Z$ nuclei $^{16}$O, $^{40}$Ca practically all isoscalar dipole strength below particle threshold is exhausted by one state, 
which coincides with the well-documented (see \cite{PPR2011} and Refs. therein) isospin-forbidden $E1$ transition. 
The electroexcitation form factor of the latter has been measured and, in the case of $^{16}$O, the charge transition density has been extracted (MIT-Bates, Ref.~\cite{But1986}). 
The result is shown in Fig.~\ref{Fig:IS-LED}(b) along with predictions within RPA~\cite{PPR2011} and a collective model~\cite{HaD1981}. 
All results point consistently to a transition density with a node close to the nuclear surface. 
The neutron and total transition density (not shown), as calculated within RPA, have almost the same profile as the proton transition density. 
The picture persists in $^{40}$Ca, $^{56}$Ni, and $^{100}$Sn. 
A small dissimilarity between the neutron and proton transition densities, as a result of the Coulomb force, gives rise to a non-vanishing $E1$ strength. 
The heavier the $N=Z$ nucleus, the stronger the effect. 
Calculations of the corresponding velocity fields have revealed an intuitive semi-classical visualization of the collective mode's dynamics: 
An oscillation of a surface layer against a denser core. 
As expected at such low energies, the oscillation is not compressional, but is effectuated by a torus in the vicinity of the surface.  

A question which arises is what happens to the IS-LED vibration as we add neutrons to the $N=Z$ core. 
Does the mode disappear? 
Is there a critical number of excess neutrons required for it to disappear? 
Does a neutron-skin oscillation appear in its place? 
If so, how many neutrons would that take? 
Refs.~\cite{PHP2012,PHP2014} set out to provide answers and the results are summarized below. 

\subsection{Ca isotopes: Strong IS-LED trasnition confirmed in $^{48}$Ca} 
In Ref.~\cite{PHP2014} a highly non-trivial prediction was made, that a strong, almost pure isoscalar dipole resonance should persist up to  
the very asymmetric nucleus $^{48}$Ca, 
below particle threshold. 
The presence of such a mode at $7.6$~MeV in $^{48}$Ca 
was subsequently confirmed in a comparison between $(\alpha ,\alpha'\gamma )$ and $(\gamma ,\gamma')$ experiments~\cite{Der2014}. 
Interestingly, the comparison did not reveal an energetic isospin splitting of $E1$ strength familiar from studies of heavier nuclei~\cite{SAZ2013}. 
Rather, largely isovector transitions lie on both sides of the isoscalar mode--one of them almost purely isovector at close proximity. 

\subsection{Sn isotopes: $^{132}$Sn not exotic enough?} 

The dipole strength in Sn isotopes, below the neutron emmission threshold,
was studied theoretically in Ref.~\cite{PHP2014}.  
The results are in line with the observed isospin splitting of $E1$ strength, with the IS-LED accounting for the isoscalar strength. 
Comparisons with available data provided illuminating insights 
and led to further quantitative predictions in the isoscalar and the $E1$ sectors--e.g., that the summed $E1$ strength below threshold shall be of 
comparable 
magnitude for $A\approx 114-132$. 
The IS-LED mode was found to persist as such up to $^{132}$Sn, i.e., its properties vary smoothly up to that neutron-rich nucleus and the protons 
still contribute to the transition density both in the nuclear interior and on the surface, even though the neutrons contribute more and more strongly--cf. the 
proton transition density in $^{132}$Sn in Fig.~\ref{Fig:IS-LED}(b). 
A genuine structural change is predicted to take effect beyond the shell closure of $^{132}$Sn, with lower-energy transitions 
dominated by loosely bound neutrons. 

\section{Where is the neutron-skin oscillation?} 

The above studies have revealed that the presence of excess neutrons can affect moderately the structure of the IS-LED mode and supply it with some $E1$ strength, 
but it is not at all instrumental in the mode's generation. 
For dramatic structural changes to the {\em lowest-energy} dipole states because of extra neutrons, 
one has to look beyond the shell closures of $^{48}$Ca and $^{132}$Sn--a 
prediction in concordance with Skyrme-based studies~\cite{ENI2013}. 
It was also seen that most of the $E1$ strength below threshold is carried by transitions {\em other than} the IS-LED--transitions which may or may not be of single-particle nature, 
and may largely lie close to the neutron-emission threshold. 

We conclude that, if there is a {\em collective} neutron-skin oscillation, 
this is not to be found in ordinary nuclei (including $^{132}$Sn, pending empirical confirmation) well below threshold, 
but either closer to threshold~\cite{Rye2002,Adr2005,Wie2009} or in much more exotic nuclei (or both). 
This conclusion certainly 
allows the resonances observed in $^{130,132}$Sn~\cite{Adr2005}, $^{68}$Ni~\cite{Wie2009} {\em above threshold} to be attributed to a neutron skin. 

\section{Summary and outlook} 

A strong isoscalar dipole resonance is known to be excited in a variety of nuclei, including isospin symmetric ones, at approximately 7~MeV. 
Based on microscopic calculations and existing data, 
the resonance is interpreted as an elementary surface vibration.  
How such a strong low-lying mode (but well above the $3^-$ or $2^+$ ones) can affect low-energy reactions is under investigation. 

The very different portions of the respective energy-weighted sum rules exhausted by the isoscalar and the isovector (or $E1$) transitions in 
isotopes studied experimentally point to very different mechanisms of generating isoscalar and $E1$ transitions, despite their energetic proximity.  
The IS-LED vibration can account for the observed isoscalar segment of pygmy dipole strength. 

It is concluded that 
genuine neutron-skin oscillations, if they exist, 
are not to be found in ordinary nuclei below threshold, but close to particle threshold~\cite{Rye2002,Adr2005,Wie2009} or in very exotic nuclei. 

Measurements on $^{132}$Sn should be able to confirm or refute 
the suggested properties of the IS-LED and a modest amount of bound low-energy $E1$ strength in this nucleus. 
A future comparison between the dipole ($E1$ and isoscalar) spectra of, e.g., $^{132}$Sn and $^{134}$Sn below and around threshold would be very useful in verifying the influence of the neutron-shell closure. 
%


\section*{Acknowledgements}
I am thankful to all my collaborators for their help 
and valued interactions in this line of work. 
This work and presentation at ARIS2014  
were supported by the Rare Isotope Science Project of the Institute for Basic Science funded by the Ministry of Science, ICT and Future Planning and the National Research Foundation of Korea (2013M7A1A1075766).

\end{document}